\documentstyle[preprint,eqsecnum,aps]{revtex}

\begin{document}
\tightenlines
\draft

\title{Action and entropy of black holes
in spacetimes with cosmological constant}

\author{Rong-Gen Cai\footnote{
Electronic address: cairg@ctp.snu.ac.kr}
}
\address{Center for Theoretical Physics,
Seoul National University, Seoul, 151-742, Korea}
\author{
Jeong-Young Ji\footnote{
Electronic address: 
jyji@phyb.snu.ac.kr}
and
Kwang-Sup Soh\footnote{
Electronic address: kssoh@phya.snu.ac.kr}
}
\address{Department of Physics Education,
Seoul National University, Seoul, 151-742, Korea}

\maketitle

\begin{abstract}

In the Euclidean path integral approach, we calculate the 
actions and the entropies for the Reissner-Nordstr\"om-de Sitter
solutions.   When the temperatures of black hole and cosmological 
horizons are equal, the entropy is the sum of one-quarter 
areas of black hole and cosmological horizons; when the inner and
outer black hole horizons coincide, the entropy is only one-quarter 
area of cosmological horizon; and the entropy vanishes when the two 
black hole horizons and cosmological horizon coincide.
We also calculate the Euler numbers of the corresponding Euclidean
manifolds, and discuss the relationship between the entropy of instanton 
and the Euler number.
 
\end{abstract}


\section{Introduction}

No doubt the discovery of Hawking~\cite{haw1} that a black hole emits 
particles as a blackbody is one of the most important achievements of 
quantum field theory in  curved spacetimes.  Due to the Hawking 
evaporation, classical general relativity, statistical physics, and 
quantum field theory are connected in the quantum black hole physics. 
Therefore it is generally believed that the deep investigation of black 
hole physics would be helpful to set up a satisfactory quantum theory of 
gravitation.   

For a Schwarzschild black hole, it is well known that the hole has a 
Hawking temperature  
\begin{equation}
T=\frac{\kappa}{2\pi}=\frac{1}{8\pi M},
\label{e1}
\end{equation}
where $\kappa$ is the surface gravity at the horizon. Shortly after the 
Hawking's discovery, Hartle and Hawking~\cite{har} rederived the black hole 
radiance in the path integral formalism. Gibbons and Perry~\cite{gib1} found 
that the thermal Green function in the black hole background has an 
intrinsic period $\beta = 2 \pi \kappa^{-1}$, which is just a 
characteristic feature of finite temperature quantum fields in  flat 
spacetimes.

Before the Hawking's discovery, Bekenstein~\cite{bek} already suggested 
that a black hole should have an entropy proportional to the area of black 
hole horizon. The work of Hawking made the entropy quantitative:
\begin{equation}
S_{\rm BH}=\frac{1}{4}A_{\rm BH},
\label{e2}
\end{equation}
where $A_{\rm BH}$ is the area of horizon.  In addition, Gibbons and 
Hawking~\cite{gib2} found that the Hawking evaporation also occurs  at the 
cosmological horizon in the de Sitter space. Further they 
observed~\cite{gib3} that the cosmological horizon has an associated entropy 
obeying the area formula:
\begin{equation}
S_{\rm CH}=\frac{1}{4}A_{\rm CH},
\label{e3}
\end{equation}
where $A_{\rm CH}$ is the area of cosmological horizon. 

When the Schwarzschild black hole is embedded into the de Sitter space, 
one has the Schwarzschild-de Sitter black hole: 
\begin{equation}
ds^2=-\left(1-\frac{2M}{r}-\frac{1}{3}\Lambda r^2\right ) dt^2
    +\left (1-\frac{2M}{r}-\frac{1}{3}\Lambda r^2\right)^{-1}dr^2
    + r^2 d \Omega ^2_2,
\label{e4}
\end{equation}
where $\Lambda$ is the cosmological constant. 
As $9 M^2\Lambda <1$, we have a black hole horizon 
$r_{\rm B}$ and a cosmological horizon 
$r_{\rm C}$ $(r_{\rm C}>r_{\rm B})$. Naturally there exist two different 
temperatures associated with the black hole horizon and cosmological 
horizon, respectively.
Combining Eqs.~(\ref{e2}) and (\ref{e3}), most people believe that the 
gravitational entropy of Schwarzschild-de Sitter black hole should be 
\begin{equation}
S=S_{\rm BH}+S_{\rm CH}=
\frac{1}{4}A_{\rm BH}+\frac{1}{4}A_{\rm CH}.
\label{e5}
\end{equation}
But the derivation of this entropy formula (\ref{e5}) is not yet 
developed. A good method to get the area 
formula of black hole entropy is the formalism of Gibbons and 
Hawking~\cite{gib3} in the Euclidean path integral method of quantum gravity 
theory. In this formalism the regular gravitational instantons (regular 
solutions of Euclidean Einstein equations) play a 
crucial role. However, this formalism cannot apply to the 
Schwarzschild-de Sitter black holes because one cannot obtain a regular 
gravitational instanton solutions in this case. By analytically
continuing Eq.~(\ref{e4}) to its Euclidean section, one can employ the 
same procedure 
used in the Schwarzschild black hole to remove the conical singularity at 
the black hole horizon or at the cosmological horizon. But one cannot 
remove the two singularities simultaneously because the temperatures 
usually do not equal to each other. Thus the Euclidean manifold is left 
with a conical singularity. An exception is the Nariai spacetime, 
which can be regarded as the limiting case of the Schwarzschild-de Sitter 
black hole as the black hole horizon and cosmological horizon coincide 
$(9 M^2 \Lambda =1)$. Its Euclidean spacetime represents a regular 
gravitational instanton with topological structure 
$S^2\times S^2$, and its metric is 
\begin{equation}
ds^2=(1-\Lambda r^2)d\tau ^2 +(1-\Lambda r^2)^{-1}dr^2 
    +\Lambda ^{-1}d\Omega ^2_2.
\label{e6}
\end{equation}
After a coordinate transformation, one can clearly see its topological
structure. The metric (\ref{e6}) can be rewritten as
\begin{equation}
ds^2=\Lambda ^{-1}(d\xi ^2 +\sin ^2 \xi d\psi ^2) 
     +\Lambda ^{-1}d\Omega ^2_2,
\label{e7}
\end{equation}
where $0 \le \psi \le 2\pi$, and $0 \le \xi \le \pi$.

The geometric features of black hole temperature and entropy seem to 
strongly imply that the black hole thermodynamics is closely related to  
nontrivial topological structure of spacetime. In recent years, there has 
been  considerable interest in this aspect of black hole physics. 
Ba\~nados, Teitelboim, and Zanelli~\cite{btz} showed that in the Euclidean 
black hole manifold with topology $R^2\times S^{d-2}$, the deficit angle 
of a cusp at any point in $R^2$ and the area of the $S^{d-2}$ are 
canonical conjugates. The black hole entropy emerges as the Euler class of 
a small disk centered at  the horizon multiplied by the area of the 
$S^{d-2}$ there. Hawking, Horowitz, and Ross~\cite{haw2} argued that, due to 
the very different topologies between the extremal black holes and 
nonextremal black holes, the area formula of entropy fails for extremal 
black holes and the entropy of extremal black holes should vanish, despite 
the non-zero area of horizon. In the Euclidean section of extremal black 
holes since there does not exist any conical singularity at the black hole 
horizon, the Euclidean time $\tau$ can 
have {\it any} period.  
Teitelboim~\cite{teit} further confirmed the zero entropy in the Hamiltonian 
formalism for extremal black holes. He put forward that the vanishing 
entropy is due to the vanishing Euler characteristic $\chi$ for extremal 
black holes. In Ref.~\cite{gib4}, Gibbons and Kallosh investigated the 
relation between the entropy and Euler number for dilaton black holes in 
some detail. They found that for extremal dilaton black holes an inner 
boundary should be introduced in addition to the outer boundary to obtain 
an integer value of the Euler number. Thus the resulting manifolds have 
(if one identifies the Euclidean time) a topology $S^1\times R \times S^2$ 
and the Euler number $\chi =0$. For the nonextremal black holes the 
topology is $R^2\times S^2$ and the Euler number $\chi=2$. More recently, 
Liberati and Pollifrone~\cite{lib} have further discussed the relation 
between the black hole entropy and the Euler number and suggested the 
following formula
\begin{equation}
S=\frac{\chi A}{8}.
\label{e8}
\end{equation}
They have checked this formula for a wide class of gravitational 
instantons such as Schwarzschild instanton, dilaton $U(1)$ black holes, 
de Sitter instanton, Nariai instanton and Kerr black holes.

In this paper, we would like to provide some evidence in favor of the 
formula (\ref{e5}) in the static, spherically symmetric
Reissner-Norstr\"om (RN)-de Sitter black holes, which are solutions of 
the Einstein-Maxwell equations  with a cosmological 
constant. When $M = |Q|$,  the temperature of 
black hole horizon is equal to the one of cosmological horizon, we have
the so-called lukewarm black hole. Since its 
analytic continuation to the Euclidean section provides a regular 
instanton, we can check the formula (\ref{e5}) in the standard Euclidean 
quantum theory of gravitation. In addition since we have two different 
horizons for this regular instanton in contrast to the Nariai instanton, 
it is of interest to see how to modify the formula (\ref{e8}) in 
this case.

The plan of this paper is as  follows. In Sec. II we first introduce 
the lukewarm black hole and the regular instanton, and then calculate the 
action and the entropy of this instanton. In Sec. III we investigate two  
special cases of RN-de Sitter solutions. One is the case where the inner
and outer horizons of black holes coincide. The solution is called a
cold black hole. The other is the so-called ultracold  solution where the
two black hole horizons and the cosmological horizon coincide. In Sec. IV we 
compute the Euler number  for lukewarm black holes, cold black holes and 
ultracold solutions, respectively, and discuss the relation between our 
results and Euler numbers. The conclusions and discussions are included 
in Sec. V.

\section{entropy of lukewarm black holes}

The action of the Einstein-Maxwell theory with a cosmological constant
is  
\begin{equation}
I=\frac{1}{16\pi}\int_{V} d^4x\sqrt{-g}
 (R-2\Lambda -F_{\mu\nu}F^{\mu\nu})
+\frac{1}{8\pi}\int_{\partial V} d^4x\sqrt{-h}K,
\label{e9}
\end{equation}
where $R$ denotes the scalar curvature, $K$ is the trace of the second 
fundamental form of surface $\partial V$,  $\Lambda$ is the cosmological 
constant and $F_{\mu\nu}$ is the Maxwell field. In the action (\ref{e9}) 
one has the static, spherically symmetric RN-de Sitter solutions (in this 
paper we consider the case of $\Lambda >0$ only, for the case $\Lambda <0 
$ results will be trivial): 
\begin{equation}
ds^2=-N^2(r)dt^2+N^{-2}(r)dr^2 +r^2d \Omega ^2_2,
\label{e10}
\end{equation}
where
\begin{equation}
N^{2}(r)=1-\frac{2M}{r} +\frac{Q^2}{r^2}-\frac{1}{3}\Lambda r^2,
\label{e11}
\end{equation}
$M$ and $Q$ are the mass and the electric charge of the solution, 
respectively, $d\Omega ^2_2$ stands for the line element on a unit 
2-sphere.  Usually the equation $N^2(r)=0$ has four roots, three positive 
real roots and a negative real root. The maximal positive root $r_{\rm C}$ 
is the cosmological horizon; the minimal ($r_{\rm A}$) is the inner 
(Cauchy) horizon of black hole; and the intermediate ($r_{\rm B}$) the 
outer horizon of black hole. Their surface gravities are
\begin{equation}
\kappa_{\rm a} =\frac{1}{2}|[N^2(r)]'|_{r=r_{\rm a}}, \  \ 
{\rm a=A,B,C}
\label{e12}
\end{equation}
where a prime denotes the derivative with respect to $r$. Thus the 
outer horizon of black hole has a  Hawking  temperature $T_{\rm B}=\kappa 
_{\rm B}/2\pi$, and the cosmological horizon also has a temperature 
$T_{\rm C}=\kappa _{\rm C}/2\pi$ different from $T_{\rm B}$. Hence this 
spacetime is unstable quantum mechanically. In its Euclidean section 
\begin{equation}
ds^2=N^2(r)d\tau ^2 +N^{-2}(r)dr^2 +r^2d \Omega ^2_2,
\label{e13}
\end{equation}
one has no way to remove simultaneously the two conical singularities at 
the black hole horizon and cosmological horizon. However, Mellor and 
Moss~\cite{mel1}, and Romans~\cite{rom} found that when $M^2=Q^2$ in 
Eq.~(\ref{e11}), one has the lukewarm RN solution where 
\begin{equation}
T_{\rm B}=T_{\rm C}=\frac{1}{2\pi}\left[\frac{\Lambda}{3}\left (1-
         4M\sqrt{\frac{\Lambda}{3}}\right)\right]^{1/2}.
\label{e14}
\end{equation}
In this case the spacetime is in thermal equilibrium with a common 
temperature $T_{\rm C}=T_{\rm B}$, and is stable classically and quantum 
mechanically. This equality (\ref{e14}) provides us a regular instanton in 
the Euclidean version (\ref{e13}), because we can remove simultaneously 
two conical singularities by requiring the imaginary time $\tau$ has a 
period $\beta =T^{-1}_{\rm C}=T^{-1}_{\rm B}$. The resulting manifold 
(\ref{e13}) has a topology $S^2 \times S^2$, which is the same as that of 
the Nariai instanton.

Following the standard method of Gibbons and Hawking~\cite{gib3}, we now 
evaluate the Euclidean action and the entropy for this lukewarm instanton 
solution. The Euclidean action can be obtained by continuing (\ref{e9}) to 
its Euclidean counterpart
\begin{equation}
I_E=-\frac{1}{16\pi}\int_{V} d^4x\sqrt{g}(R-2\Lambda 
   -F_{\mu\nu}F^{\mu\nu})-\frac{1}{8\pi}\int _{\partial V}
   d^3x \sqrt{h} (K-K_0),
\label{e15}
\end{equation}
where we have introduced a subtraction term $K_0$, which is the
trace of the second fundamental form of the reference background. In the 
case of lukewarm solution, we do not have to consider the boundary term in 
the action (\ref{e15}), because the topology is $S^2\times S^2$ which has 
no boundary. Using the Einstein equation
\begin{equation}
R_{\mu\nu}-\frac{1}{2}R g_{\mu\nu} +\Lambda g_{\mu\nu}=
        2\left(F_{\mu\rho}F^{\rho}_{~\nu}-\frac{1}{4}
          g_{\mu\nu}  F^2\right),
\label{e16}
\end{equation}
and the scalar curvature 
\begin{equation}
R=-g^{-1/2}[g^{1/2}(N^2)']'-2G_{~0}^{0},
\label{e17}
\end{equation}
where $G^0_{~0}$ is the 0-0 component of the Einstein tensor, we can 
easily obtain
\begin{equation}
I_E=-\pi r_{\rm C}^2 -\pi r_{\rm B}^2 -\beta Q^2
\left(\frac{1}{r_{\rm B}}-\frac{1}{r_{\rm C}}\right).
\label{e18}
\end{equation}
It is instructive to compare the action (\ref{e18}) with the one of RN 
black holes with topology $R^2\times S^2$ in the grand canonical 
ensemble~\cite{bard}. In the latter $\beta M$ replaces the first term in 
(\ref{e18}). So different topological structures would result in very 
different Euclidean actions. In the action (\ref{e18}), the first term is 
obviously one quarter of the area of cosmological horizon; the second one 
is one quarter of the area of black hole horizon; and the last one is the 
difference of electric potential energy between at the cosmological and 
the black hole horizons. When $Q=0$ the action (\ref{e18}) also includes 
two special cases in Ref.\cite{gib3}: For the de Sitter space $(M=0)$ with 
$r_{\rm C} =\sqrt{\Lambda / 3}$, the first term gives the same result; 
for the Nariai instanton, 
with $r_{\rm B}=r_{\rm C}=1/\sqrt{\Lambda}=3M$, 
first two terms give the same result.

In order to get the entropy of the instanton, it is helpful to employ the 
general argument given by Kallosh, Ort\'\i n, and Peet~\cite{kal}. In a 
thermodynamic system with conserved charges $C_i$ and corresponding  
chemical potentials $\mu _i$, the starting point to study this system is 
the grand partition function ${\cal Z}$ in a grand canonical ensemble 
\begin{equation}
{\cal Z}={\rm Tr} e^{-\beta (H-\mu _i C_i)}. 
\label{e19}
\end{equation}
The thermodynamic potential $W=E-TS -\mu _i C_i$ can be obtained from the 
partition function as
\begin{equation}
W=-T \ln {\cal Z},
\label{e20}
\end{equation}
from which the entropy $S$ is
\begin{equation}
S=\beta (E-\mu _i C_i) + \ln {\cal Z}.
\label{e21}
\end{equation}
In the Euclidean quantum theory of gravity, the partition function 
is~\cite{gib3}
\begin{equation}
{\cal Z}=\int D[\phi]D[g]\exp (-I_E[g, \phi]),
\label{e22}
\end{equation}
where $\phi$ represents the matter fields and $I_E$ is the Euclidean 
action. Naturally it is expected that the dominant 
contribution  of the integration in (\ref{e22}) comes from the field 
configurations satisfying the classical Euclidean field equations.
Under the zero-loop approximation, Gibbons and Hawking \cite{gib3}
have shown that the partition function is
\begin{equation}
{\cal Z}=\exp[-I_E(g_0,\phi _0)],
\label{e23}
\end{equation}
where $I_E(g_0, \phi _0)$  denotes the on-shell Euclidean action of the 
classical instanton. For asymptotically flat or anti-de Sitter nonextremal 
black holes, the manifold of instantons has only a boundary at the spatial 
infinity and is regular at the black hole horizon. For extremal black 
holes, Gibbons and Kallosh~\cite{gib4}, and Hawking {\it et
al.}~\cite{haw2} 
have argued that the manifold has an inner boundary corresponding to the 
black hole horizon, in addition to the outer boundary. In our case the 
manifold is regular both at the black hole horizon and the cosmological 
horizon. However, for the later use, we label the on-shell action 
$I_E(g_0, \phi _0)$ as
\begin{equation}
I_E(g_0, \phi _0)=I_E|^{r_{\rm out}}_{r_{\rm in}},
\label{e24}
\end{equation}
where $r_{\rm out}$ and $r_{\rm in}$ represent the outer and inner 
boundaries, respectively.  To get the entropy from (\ref{e21}), one must 
calculate the quantity $E-\mu _i C_i$ by considering the following 
amplitude for imaginary time between two surfaces of Euclidean times $\tau 
_1$ and $\tau _2$ with given boundary conditions~\cite{gib3,kal}:
\begin{equation}
\langle \tau _1|\tau _2\rangle =
e^{-(\tau _2-\tau _1)(E-\mu _i C_i)}.
\label{e27}
\end{equation}
From the above equation,  one has
\begin{equation}
\beta (E-\mu _i C_i)=I_E|^{R_{\rm out}}_{R_{\rm in} }
\label{e28}
\end{equation}
for $\tau_2-\tau_1 =\beta$. Here $R_{\rm in}$ and $R_{\rm out}$ stand for 
the inner and outer physical boundaries of the spacetime. They are the 
black hole horizon and cosmological horizon for the lukewarm black hole 
case, respectively.

The two Euclidean actions can be expressed as
\begin{eqnarray}
&& I_E|^{R_{\rm out}}_{R_{\rm in}}=-\frac{1}{16\pi} \int_{V}
           d^4x \sqrt{g} (R + {\cal L}_{\rm matter})
       -\frac{1}{8\pi}\int^{R_{\rm out}}_{R_{\rm in}}
           d^3x\sqrt{h}
         (K-K_0),
\label{e29} \\
&&I_E|^{r_{\rm out}}_{r_{\rm in}}=-\frac{1}{16\pi} \int _{V}
         d^4x \sqrt{g} (R + {\cal L}_{\rm matter}) 
      -\frac{1}{8\pi}\int ^{r_{\rm out}}_{r_{\rm in}}
        d^3x \sqrt{h}
         (K-K_0), 
\label{e30}
\end{eqnarray}
where ${\cal L}_{\rm matter}$ is the Lagrangian for matter fields. 
Substituting (\ref{e29}) and (\ref{e30}) into (\ref{e21}), one has 
\begin{eqnarray}
S & =& I_E|^{R_{\rm out}}_{R_{\rm in}}
        -I_E|^{r_{\rm out}}_{r_{\rm in}} \nonumber\\
  &=& -\frac{1}{8\pi}\int ^{R_{\rm out}}_{R_{\rm in}}d^3x\sqrt{h}
      (K -K_0) +\frac{1}{8\pi} \int ^{r_{\rm out}}_{r_{\rm in}} 
   d^3x\sqrt{h}(K -K_0).
\label{e31}
\end{eqnarray}
As we have mentioned above, the Euclidean manifold $S^2 \times S^2$ has 
no boundary, consequently the second term in (\ref{e31}) can be dropped 
out. The extrinsic curvature $K$ in the metric (\ref{e13}) for a timelike 
surface fixed 
$r~(r_{\rm B} <r<r_{\rm C})$ is
\begin{equation}
K=-g^{-1/2}(N g^{1/2})'.
\label{e32}
\end{equation}
Now we choose $K_0$ so that the boundary become asymptotically imbeddable 
as one goes to larger and larger radii in an asymptotically de Sitter 
space:
\begin{equation}
K_0=- r^{-2}(r^2\sqrt{1-\Lambda r^2/3})'.
\label{e33}
\end{equation}
Substituting Eqs.~(\ref{e32}) and (\ref{e33}) into (\ref{e31}), we obtain
\begin{eqnarray}
S&=&\pi r_{\rm C}^2 +\pi r_{\rm B}^2 \nonumber \\
 &=&\frac{A_{\rm CH}}{4}+\frac{A_{\rm BH}}{4}.
\label{e34}
\end{eqnarray}
Thus we get indeed the entropy formula (\ref{e5}) in the lukewarm black 
hole model. Comparing (\ref{e34}) with (\ref{e18}), we can clearly see 
that the Euclidean action is no longer equal to the entropy of instanton. 
This is very different from that of asymptotically flat black holes, where  
they are always equal to each other~\cite{gib3,kal}. In  fact, the situation 
in which they are not equal already appears in the
de Sitter space and the Nariai instanton~\cite{gib3}.

\section{Cold black holes and ultracold solutions}

In Ref.~\cite{rom} Romans have classified in detail the static,
spherically symmetric  
solutions of the Einstein-Maxwell equations with a cosmological constant. 
In addition to the lukewarm black hole discussed in the previous section, 
there exist two kinds of solutions of interest, cold black holes and 
ultracold solutions.
In this section, we will discuss them separately.

\subsection{Cold black holes}

When the inner and outer horizons of the RN-de Sitter black hole 
coincide with each other, it is called a cold black hole, which 
corresponds to the extremal black hole in the asymptotically flat or 
anti-de Sitter spacetime. In this case, the temperature of black hole 
horizon vanishes, but the cosmological horizon has still a nonvanishing 
Hawking temperature. Since its Euclidean section has only a conical 
singularity at the cosmological horizon, we can remove it by identifying 
the imaginary time with the period
\begin{equation}
\beta =T_{\rm C}^{-1},
\label{e35}
\end{equation}
where $T_{\rm C}$ is the Hawking temperature of the cosmological horizon. 
Then the resulting manifold has an inner boundary at the black hole 
horizon ($r_{\rm B}$). The Euclidean action becomes
\begin{equation}
I_E=-\pi r_{\rm C}^2 -\beta Q^2 \left (\frac{1}{r_{\rm B}}-
           \frac{1}{r_{\rm C}}\right ),
\label{e36}
\end{equation}
The second term in Eq.~(\ref{e18}) now is canceled out by the surface 
term coming from  the black hole horizon in Eq. (\ref{e15}). In the 
entropy formula (\ref{e31}), the inner boundary $r_{\rm in}$ is the same 
as the boundary $R_{\rm in}$, that is, they are both the black hole 
horizon. $R_{\rm out}$ is the cosmological horizon and $r_{\rm out}$ is 
absent. From (\ref{e31}) it follows directly that
\begin{equation}
S=\pi r_{\rm C}^2.
\label{e37}
\end{equation}
Obviously, the entropy has only the contribution from the cosmological 
horizon.

\subsection{Ultracold solutions}

When the inner and outer horizons and the cosmological horizon coincide, 
that is the equation $N^2(r)=0$ has three same positive roots, this RN-de 
Sitter solution is called an ultracold one. Then the  physical region is 
$0 \le r\le r_{\rm C}$, where $r_{\rm C}$ denotes the triple root. The 
situation is somewhat similar to the de Sitter space. But there are 
differences:
\begin{equation}
N^2(r)|_{r=r_{\rm C}}=0, \ \ \ [ N^2(r)]'|_{r=r_{\rm C}}\ne 0
\label{e38}
\end{equation}
for the de Sitter space; 
\begin{equation}
N^2(r)|_{r=r_{\rm C}}=[N^2(r)]'|_{r=r_{\rm C}}=[N^2(r)]''|_{r
=r_{\rm C}}=0,
\ \ [N^2(r)]'''|_{r=r_{\rm C}} < 0
\label{e39}
\end{equation}
for the ultracold solution. In addition, another important point is that 
$r=0$ is a naked singularity for the ultracold solutions. Similar to the 
extremal black holes, the Euclidean section of ultracold solution is 
regular at $r_{\rm C}$, we can identify the Euclidean time with {\it any} 
period. In this case the outer boundary $R_{\rm out}$ is $r_{\rm C}$. 
Similar to the extremal black holes in asymptotically flat or anti-de 
Sitter spacetimes, we must introduce another outer boundary $r_{\rm out}$ 
at $r_{\rm C}$ in the Euclidean manifold. Furthermore, to remove the naked 
singularity at $r=0$, we introduce an inner boundary $R_{\rm in}=r_{\rm 
in}=\varepsilon$, where $\varepsilon $ is a small positive quantity. Thus, 
from Eq.~(\ref{e31}) we have \begin{equation}
S=0,
\label{e40}
\end{equation}
for the ultracold solutions. The Euclidean action, from (\ref{e15}) and 
(\ref{e30}), is
\begin{equation}
I_E=-\beta Q^2\left (\frac{1}{\varepsilon}-
\frac{1}{r_{\rm C}}\right),
\label{e41}
\end{equation}
where $\beta $ is the period of the Euclidean time $\tau$. From 
(\ref{e41}) we can see that the action diverges as $\varepsilon 
\rightarrow 0$. This reflects the fact that the electric potential energy 
is divergent for a point-like charge.

\section {Euler numbers and entropy of instantons}

In the previous sections we have obtained that the entropy of lukewarm 
black holes is one quarter of the sum of the areas of the black hole 
horizon and the cosmological horizon, the entropy of cold black hole is 
only one quarter of the area of cosmological horizon, and the entropy of 
the ultracold solutions vanishes. Evidently the formula (\ref{e8}) of 
Liberati and Pollifrone cannot apply to our results. In order to explain 
these results it is instructive to investigate the topological properties 
of manifolds. For the lukewarm black hole, its topological structure is 
$S^2 \times S^2$. So the Euler number of the manifold is $\chi =4$. For 
our purposes, however, it is helpful to reexamine the Gauss-Bonnet 
integral.

In the four dimensional Riemannian manifold, the Gauss-Bonnet integral 
is~\cite{rep}
\begin{equation}
S_{\rm GB}^{\rm volume}=\frac{1}{32\pi ^2}\int _{V}
\varepsilon _{abcd} R^{ab}\wedge R^{cd},
\label{e42}
\end{equation}
where $R^{ab}$ is the curvature two-form defined as
\begin{equation}
R^a_{\ b}=d\omega ^a_{\ b} +\omega ^a_{\ c}
          \wedge \omega ^{c}_{\ b},
\label{e43}
\end{equation}
and $\omega ^{a}_{\ b}$ is the spin connection one-form. For a closed 
Riemannian manifold without  boundary, its Euler number $\chi$ is given by 
the Gauss-Bonnet integral (\ref{e42}). For a manifold with boundary, the 
exact Euler number is obtained by adding a term integrated over the 
boundary to Eq. (\ref{e42}), that is
\begin{equation}
\chi =S_{\rm GB}^{\rm volume} + S_{\rm GB}^{\rm boundary},
\label{e44}
\end{equation}
where
\begin{equation}
S_{\rm GB}^{\rm boundary}=-\frac{1}{32\pi ^2}\int _{\partial V}
       \varepsilon_{abcd}(2\theta ^{ab}\wedge R^{cd}
       -\frac{4}{3}\theta ^{ab}\wedge \theta ^{c}_{\ e}
      \wedge \theta ^{ed}),
\label{e45}
\end{equation}
where $\theta^{ab}$ is the second fundamental form of the boundary
$\partial V$.  In the static, spherically symmetric solution (\ref{e13}), 
the volume 
integral of Gauss-Bonnet action is~\cite{gib4}
\begin{eqnarray}
S_{\rm GB}^{\rm volume}&=&-\frac{\beta}{2\pi}
\int ^{R_{\rm out}}_{R_{\rm in}} dr \frac{\partial}
{\partial r}[(N^2)'(1-N^2)] \nonumber \\
&=&\frac{\beta}{2\pi}\{[(N^2)'(1-N^2)]|_{r=R_{\rm in}}
-[(N^2)'(1-N^2)]|_{r=R_{\rm out}}\},
\label{e46}
\end{eqnarray}
where $\beta$ is the period of the Euclidean time $\tau$, and the 
boundary term (\ref{e45}) is
\begin{eqnarray}
S_{\rm GB}^{\rm boundary}&=&\frac{\beta}{2\pi}
[(N^2)'(1-N^2)]|^{r_{\rm out}}_{r_{\rm in}}
\nonumber \\
&=&-\frac{\beta}{2\pi}\{[(N^2)'(1-N^2)]|_{r_{\rm in}}
-[(N^2)'(1-N^2)]|_{r_{\rm out}}\}.
\label{e47}
\end{eqnarray}

(i) For the lukewarm black holes, the topology of manifold is  $S^2 
\times S^2$, that is, it is a compact manifold without boundary. The 
Euclidean time has an intrinsic period
\begin{equation}
\beta =T^{-1}_{\rm B}=T^{-1}_{\rm C}
=4\pi [(N^2)'|_{r=r_{\rm B}}]^{-1}
=4\pi [|(N^2)'|_{r=r_{\rm C}}]^{-1},
\label{e48}
\end{equation} 
Thus the Euler number is only given by Eq.~(\ref{e46}). Substituting 
(\ref{e48}) into (\ref{e46}) yields
\begin{equation}
\chi = 2 -(-2)=2+2=4.
\label{e49}
\end{equation}
Clearly, we can understand that the first $2$ in (\ref{e49}) comes from 
the black hole horizon and the second from the cosmological horizon.

(ii) For the cold black holes, the Euclidean time has no intrinsic period 
at the black hole horizon because there is no conical singularity there, 
so the period $\beta $ can be arbitrary. But there is a conical 
singularity at the cosmological horizon, in order to remove the 
singularity, we must identify the $\tau$ with an intrinsic period
\begin{equation}
\beta =4\pi [|(N^2)'|_{r=r_{\rm C}}]^{-1}.
\label{e50}
\end{equation}
In this case, the resulting manifold has the topology $R^2\times S^2$, 
we must consider the boundary term (\ref{e47}). The outer 
boundary $r_{\rm out}$ is absent. However, we must introduce an inner 
boundary at 
$ r_{\rm in} = r_{\rm B}+\varepsilon  $, as in the extremal black holes 
in the asymptotically flat space~\cite{gib4}. Combining (\ref{e46}) and 
(\ref{e47}), we have 
\begin{equation}
\chi = 0-(-2)=2,
\label{e51}
\end{equation}
from which we can see clearly that the cosmological horizon has 
the contribution to the Euler number only.

(iii) For ultracold solutions, the Euclidean time  has no intrinsic 
period. Thus $\beta $ can be an arbitrary finite value, but 
\begin{equation}
N^2(r)=[N^2(r)]'=0,
\label{e52}
\end{equation}
at $r=r_{\rm C}$. In this case, the Euclidean manifold has the
topological structure $S^1\times R^1\times S^2$,  we must  introduce
not only the outer 
boundary $R_{\rm out}=r_{\rm out}=r_{\rm C}$ but also the inner boundary 
$R_{\rm in}=r_{\rm in}=\varepsilon$, because $N^2(r)$ and $[N^2(r)]'$ are 
both divergent at $r=0$. From (\ref{e46}) and (\ref{e47}), we obtain
\begin{equation}
\chi =0+0=0. 
\label{e53}
\end{equation}

What is the relation between the black hole entropy and the Euler number?  
The relation (\ref{e8}) of Liberati and Pollifrone does not apply to our 
case. In order to relate the entropy of instantons to the Euler numbers, 
an interesting suggestion is to divide the Euler number into two parts:
\begin{equation}
\chi =\chi_1 +\chi_2.
\label{e54}
\end{equation}
For example, we could think that $\chi_1$ is the contribution of the 
black hole horizon and $\chi _2$ comes from the cosmological horizon. From 
the calculations made above, this division seems reasonable. Thus we have 
the following relation
\begin{equation}
S=\frac{\chi _1}{8}A_{\rm BH} +\frac{\chi _2}{8}A_{\rm CH}.
\label{e55}
\end{equation}
This formula contains some known special cases.  
For asymptotically flat or anti-de Sitter nonextremal black holes, the 
cosmological horizon is absent. Thus we have $\chi _1=2$ and $\chi _2=0$. 
The outer boundary contributes a zero result, the black hole entropy is 
$S=\chi_1 A_{\rm BH}/8$. For the extremal black holes,  $\chi_1$ and $\chi 
_2$ both vanish: $ \chi _1=\chi _2=0$, therefore $S=0$. For the de Sitter 
space, the black hole horizon is absent, we have $\chi _1=0$ and $\chi 
_2=2$, and the entropy of de Sitter space is $S=\chi _2 A_{\rm CH}/8$; For 
the Nariai instanton which can be regarded as the limiting case of the 
Schwarzschild-de Sitter black holes, $\chi _1=\chi_2 =2$, and the area of 
black hole horizon equals to the one of cosmological horizon. The entropy 
of the Nariai instanton obeys the relation (\ref{e55}). For lukewarm black 
holes, cold black holes and ultracold solutions, their entropy satisfies 
manifestly the formula (\ref{e55}).

\section{Discussion}

In this paper we have calculated the action and entropy of lukewarm black 
holes, cold black holes and ultracold solutions in the Einstein-Maxwell 
theory with a cosmological constant. Our calculations have been performed 
in the formalism of the Euclidean quantum theory of gravitation under the 
zero-loop approximation. We have found that in the lukewarm black holes, 
the action and entropy are no longer equal to each other. The entropy of 
lukewarm black holes is the sum of entropies of black hole  horizon and  
cosmological horizon, which provides an evidence in favor of the entropy 
formula (\ref{e3}) of de Sitter black holes. For the cold black holes, the 
gravitational entropy is contributed by only the cosmological horizon, 
and is 
one quarter of the area of the cosmological horizon. For the ultracold 
solutions, although the spacetime is somewhat similar to the de Sitter 
space, the entropy vanishes identically, as in the case of extremal black 
holes in the asymptotically flat or anti-de Sitter spacetimes.
Further, we have investigated the 
topological properties of manifolds corresponding to the respective 
solutions and calculated their Euler numbers. In order to relate the 
entropy with the Euler number, we present an interesting relation 
(\ref{e55}), in which we divide the Euler number of manifolds into two 
parts: One comes from the black hole horizon and the other from the 
cosmological horizon. Of course, it should be stressed here that some 
results  related to the cold black hole and ultracold solution 
 strongly rely on the some arguments about extremal black holes 
 proposed in Refs.\cite{haw2,teit,gib4}.  Here it  also  should be  
noticed  that in the semiclassical canonical quantum theory of 
gravity Brotz and Kiefer~\cite{BrotzK97} obtained  that the extremal RN 
black holes have zero entropy.

Over the past two years, however, the statistical explanation of
black hole entropy in 
the string theory has shown that the extremal black hole entropy still 
obeys the area formula~\cite{stro}. How to understand the two seemingly 
contradictory conclusions obtained in the Euclidean path integral method
and in the string theory, respectively? It may be helpful to consider
the conditions under which the two different results are deduced. In the 
string theory, some extremal black holes correspond to the BPS 
saturated states, in which $M$ and $Q$ are already quantized quantities.
That is, 
that the entropy of extremal black holes still obeys the area formula in
the string theory is obtained by taking the extremality condition
$M=|Q|$ after quantization. In the Euclidean path integral method, 
the conclusion of extremal black holes having zero entropy is derived
under first fixing the topology of Euclidean manifold corresponding to 
the extremal black holes and then quantizing the theory
\cite{haw2,teit,gib4}. Just as argued recently by Ghosh and Mitra
\cite{gho}, if one does not first fix the topology and quantizes the
theory under the certain boundary conditions in the path integral 
formalism, and then takes the extremality condition for the extremal
black holes, the resulting entropy of extremal black holes satisfies 
the area formula. The entropy in fact comes from the topology of
non-extremal black holes.  Ref.\cite{gho} clearly demonstrates how the
topology of non-extremal black holes enters into the partition function
of extremal black holes. Note that the result of zero entropy for
extremal black holes in the semiclassical canonical quantum gravity is
also derived under first fixing the topological structure (extremality 
condition) of extremal black holes before the quantization
\cite{BrotzK97}. The results in string theory show that the corrections
of strings may affect drastically the geometry near the horizon of black
holes.  Therefore, the two conclusions are not in contradiction
with each other in the sense that the entropy vanishes for the fixed 
topology, extremal black holes. Thus, the entropy of
extremal black holes seems to become relevant to  one's understanding 
of  the extremal black holes. Obviously,  further
understanding for  extremal black holes is needed.

As for the general black holes in de Sitter space, an important problem 
is to develop a satisfactory method to derive the gravitational
entropy.  As for this point, the off-shell approach seems a promising 
direction.  Although the relation (\ref{e8}) or (\ref{e55}) can explain 
some known results, to what extent 
does the relation remain valid? These issues are currently under 
investigation. Finally, we would like to point out that some of our 
conclusions are also valid for lukewarm black holes in the Kerr-de Sitter 
solutions~\cite{mel2}.

\section*{Acknowledgments}

This work was supported by the Center for Theoretical Physics (CTP)
of Seoul National University and the Basic Science Research Institute 
Program, Ministry of Education Project No. BSRI-97-2418. 
R.G.C. would like to express his  sincere 
thanks to Profs. S. P. Kim, C. K. Lee, and H. S. Song 
 for their kind help and generosity, Dr. J. H. Cho  for very 
stimulating discussions. We are grateful to R. Mann for informing us
of the reference \cite{Mann}, in which the action of some black holes
considered in the present paper. has also been calculated in
the context of the pair production of black holes.

\end{document}